\documentclass[a4paper]{jpconf}
\usepackage{graphicx}
\begin{document}
\title{Calorimetric study of the superconducting and normal state properties of Ca(Fe$_{1-x}$Co$_{x}$)$_{2}$As$_{2}$}

\author{M.~Abdel-Hafiez$^{1}$, L. Harnagea$^{1}$, S. Singh$^{2}$, U. Stockert$^{3}$, S. Wurmehl$^{1}$, R. Klingeler$^{4}$, A. U. B.~Wolter$^{1}$, B. B\"uchner$^{1}$}

\address{$^{1}$ Leibniz-Institute for Solid State and Materials Research, D-01171 Dresden, Germany}
\address{$^{2}$ Indian Institute of Science Education and Research (IISER), Maharashtra 411008, India}
\address{$^{3}$ MPI for Chemical Physics of Solids, D-01187 Dresden, Germany}
\address{$^{4}$ Kirchhoff Institute for Physics, University of Heidelberg, D-69120 Heidelberg, Germany}
\ead{m.mohamed@ifw-dresden.de}

\begin{abstract}
We present a calorimetric study on single crystals of
Ca(Fe$_{1-\emph{x}}$Co$_{\emph{x}}$)$_{2}$As$_{2}$ (\emph{x} = 0, 0.032, 0.051, 0.056,
0.063, and 0.146). The combined first order spin-density wave/structural transition occurs in the parent CaFe$_{2}$As$_{2}$ compound
at 168 K and gradually shifts to lower
temperature for low doping levels (\emph{x} = 0.032 and \emph{x} = 0.051). It is
completely suppressed upon higher doping \emph{x} $\geq$ 0.056.
Simultaneously, superconductivity appears at lower temperature
with a transition temperature around \emph{T$_{c}$} $\sim$ 14.1 K for
Ca(Fe$_{0.937}$Co$_{0.063}$)$_{2}$As$_{2}$. The phase diagram of
Ca(Fe$_{0.937}$Co$_{0.063}$)$_{2}$As$_{2}$ has been derived and
the upper critical field is found to be $\mu$$_{0}$$\emph{H}_{c2}^{(\emph{c})} = 11.5$ T
and $\mu$$_{0}$$\emph{H}_{\emph{c}2}^{(\emph{ab})}=19.4$ T for the $\emph{c}$ and $\emph{ab}$ directions,
respectively.
\end{abstract}

\section{Introduction}

After the discovery of superconductivity in LaFeAsO (1111
compound) with a transition temperature T$_{c}$ $\sim$ 26 K
\cite{Kamihara2008}, a lot of investigations have been undertaken
finding an increase of T$_{c}$ by replacing La with
smaller size rare earth ions \cite{Chen2008} reaching up to
\emph{T$_{c}$} = 55 K in SmO$_{1-x}$F$_{x}$FeAs \cite{Ren Z-A2008}.
Simultaneously, several other families of Fe-based layered
superconductors have been discovered, such as the ThCr$_2$Si$_2$
structure-type \emph{A}Fe$_2$As$_2$ ('122' family, \emph{A} = Ca, Sr, Ba, Eu).
They adopt a structure analogous to oxypnictides with LaO
layers replaced by layers of \emph{A}. These series are of great interest
because large single crystals can be grown
\cite{Harnagea2011} which show a structural transition from a
tetragonal to an orthorhombic structure combined with an
antiferromagnetic ground state due to a spin-density wave (SDW)
formation \cite{Ni2008}. Furthermore, the parent compound of
the 122 series becomes superconducting upon electron and hole
doping \cite{MR2008},\cite{NN2009}. The CaFe$_{2}$As$_{2}$ system is of particular
interest because the application of a pressure of 0.69 GPa is sufficient to suppress
the SDW/structural transition completely and to induce
superconductivity with a \emph{T$_{c}$} exceeding 10 K
\cite{Park2008},\cite{Baek2009} while for SrFe$_{2}$As$_{2}$ and
BaFe$_{2}$As$_{2}$ higher pressure up to 4 GPa are needed \cite{Alireza2009}.
We should also mention that the electron and hole doped BaFe$_{2}$As$_{2}$ and SrFe$_{2}$As$_{2}$
series of compounds have remained very much in focus, while studies on the
analogous CaFe$_{2}$As$_{2}$ compounds are very scarce.

Since specific heat measurements can provide useful information about the electronic and the
gap structural properties, it is important to explore the normal state and superconducting
properties of Ca(Fe$_{1-\emph{x}}$Co$_{\emph{x}}$)$_{2}$As$_{2}$ single crystals using this technique. In this contribution, we present a first calorimetric study of both
antiferromagnetic and superconducting
Ca(Fe$_{1-x}$Co$_{x}$)$_{2}$As$_{2}$ for (\emph{x} = 0, 0.032, 0.051,
0.056, 0.063 and 0.146) and report on the superconducting
properties of the Ca(Fe$_{0.937}$Co$_{0.063}$)$_{2}$As$_{2}$ single
crystal.

\section{Experimental}

Single crystals of Ca(Fe$_{1-x}$Co$_{x}$)$_{2}$As$_{2}$ were grown
by the high temperature solution growth method using Sn flux. All
room-temperature processing (weighing, mixing, grinding and
storage) of these materials was carried out in an Ar filled
glove-box (O$_{2}$ and moisture level less than 0.1 ppm). The crystal
growth and the physical properties of the Ca(Fe$_{1-\emph{x}}$Co$_{\emph{x}}$)$_{2}$As$_{2}$ series
 are presented in details in ref.\cite{Harnagea2011},\cite{Klingeler2010}-\cite{Prokes2011} and the \emph{x} values
are indicating the composition as measured by Energy-dispersive X-ray spectroscopy (EDX) \cite{Harnagea2011}.
The specific heat measurements have been performed in a Physical
Property Measurement System (PPMS) using a relaxation technique.
For the measurements \emph{B} $\parallel$ \emph{ab} a
small copper block has been used to mount the sample on the
specific heat puck. The heat capacity of the copper block was
determined in a separate measurement and directly subtracted.
Because of the nature of the structural transition, a temperature
rise of only 0.5~\% was used for those measurements in the
vicinity of the transition.

\section{Results and discussions}

\begin{figure}\center
\includegraphics[width=26pc,clip]{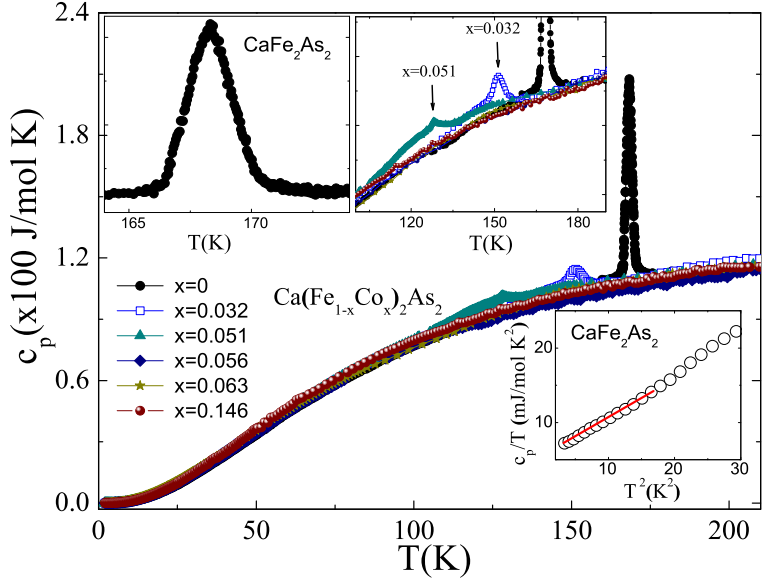}
\caption{Temperature dependence of the specific heat of
Ca(Fe$_{1-\emph{x}}$Co$_{\emph{x}}$)$_{2}$As$_{2}$ for \emph{x} = 0, 0.032, 0.051, 0.056,
0.063, and 0.146 measured in zero magnetic field. The lower inset
depicts the low temperature specific heat of the parent CaFe$_{2}$As$_{2}$ compound
 together with a fit to
\emph{c$_{p}$=$\gamma$T+$\beta$T$^{3}$} shown as a red solid line. The
upper insets show a zoom into the region around the phase
transition for CaFe$_{2}$As$_{2}$ plotted as \emph{c$_{p}$} vs. \emph{T} and upon Co doping.}
\label{Fig:3}
\end{figure}

Fig. 1 summarizes the temperature dependence of
the zero-field specific heat at various Co-dopings of
Ca(Fe$_{1-\emph{x}}$Co$_{\emph{x}}$)$_{2}$As$_{2}$ between 1.8 and 220\,K. In
the parent compound (\emph{x} = 0) a sharp first-order structural
transition coincides with a SDW transition at \emph{T$_{s}$} = 168\,K
(upon warming), which is in a good agreement with literature
\cite{Harnagea2011},\cite{KumarN2009},\cite{Ronning2008}. The transition width is
estimated to be about 4 K. To determine the Sommerfield
coefficient $\gamma$ of our CaFe$_{2}$As$_{2}$ single crystal, the
low-temperature specific heat data are plotted as \emph{$c_{p}/T$} as a
function of \emph{$T^{2}$}. These data can be fitted to a \emph{$c_{p}$=$\gamma
T$+$\beta T^{3}$} power law, where \emph{$\gamma$} and \emph{$\beta$} are determined by
the electronic and phononic contributions, respectively, as shown in
the lower inset in Fig. 1. The \emph{$\gamma$} value is found to be
around 5.4 mJ/mol K$^{2}$, which is in agreement with values
ranging between 4.7 - 8.2 mJ/mol K$^{2}$ in the literature
\cite{Ni2008},\cite{Ronning2008}. At present we do not understand this relatively
large sample dependent variation in the measured values of $\gamma$.
It is, however, conceivable that this difference is due to a different
single crystal growth technique which is leading to differences
in the crystal quality. The phononic coefficient \emph{$\beta$} is
found to be $\sim$ 0.508 mJ/mol K$^{4}$. Using the relation
\emph{$\theta_{D}$ = (12$\pi$$^{4}$RN/5$\beta$) $^{1/3}$}, where \emph{$R$} is
the molar gas constant and \emph{$N$} = 5 is the number of atoms per
formula unit, we obtain the Debye temperature \emph{$\theta_{D}$} = 267 K.

Upon Co doping of the system the combined SDW/structural
transition is shifted to lower temperature and becomes broader and
considerably reduced in magnitude, with the transition being
shifted to 148 K and 128 K for \emph{x} = 0.032 and \emph{x} = 0.051,
respectively, as shown in the upper inset in Fig. 1. Contrary to the magnetization, resistivity and neutron diffraction studies \cite{Harnagea2011},\cite{Prokes2011}, in our specific heat studies no evident splitting of the structural and magnetic transition was found in our sample \emph{x} = 0.051 and \emph{x} = 0.056. For higher Co doping the transitions are completely suppressed and superconductivity evolves at low temperature. The transitions observed in the specific heat data are in general consistent with those found in the resistivity, magnetization and neutron diffraction \cite{Harnagea2011},\cite{Klingeler2010},\cite{Prokes2011} measurements performed in few of Ca(Fe$_{1-\emph{x}}$Co$_{\emph{x}}$)$_{2}$As$_{2}$ single crystals from the same batch like those used in the present study.

\begin{figure}[t,center]\vspace{-1.0pc}
\includegraphics[width=22pc,clip]{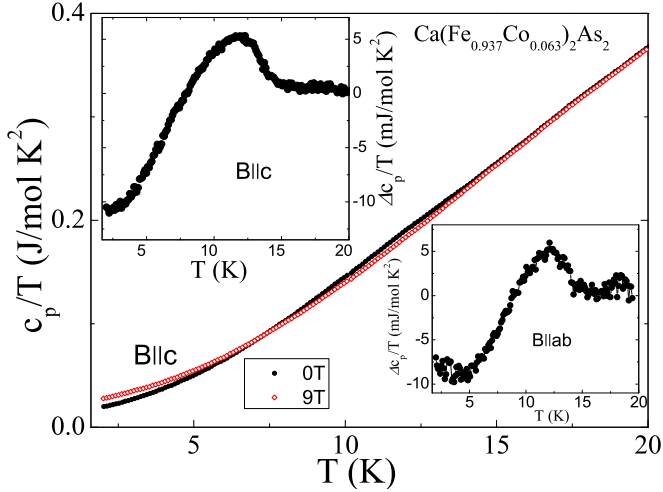}\hspace{2pc}%
\begin{minipage}[b]{15pc}\caption{\label{label}Specific heat measurements of Ca(Fe$_{0.937}$Co$_{0.063}$)$_{2}$As$_{2}$ measured in zero and 9 T for $\emph{B} \parallel \emph{c}$. The upper inset shows the difference
 between the zero and 9 T data, \emph{$\Delta c_{p}/T$}, which depicts a specific heat jump with a superconducting
transition temperature of about 14.1 K for $\emph{B} \parallel \emph{c}$. The lower inset shows the difference between the
zero and 9 T data, \emph{$\Delta c_{p}/T$}, for $\emph{B} \parallel \emph{ab}$.}\label{8}
\end{minipage}
\end{figure}

Figure 2 shows the temperature dependence of the heat capacity of
Ca(Fe$_{0.937}$Co$_{0.063}$)$_{2}$As$_{2}$ in zero and 9\,T for $\emph{B}
\parallel \emph{c}$. Taking into account the relatively
broad superconducting transition in both resistivity and
susceptibility measurements on the same single crystal as
discussed in detail in \cite{Harnagea2011}, the apparent lack of the anomaly at
\emph{T$_{c}$} may be due to the relatively large transition width.
This is strongly supported by specific-heat data on superconducting CaFe$_{1.94}$Co$_{0.06}$As$_{2}$
single crystals, where no superconducting transition has been
observed either \cite{Kumar2009}. This probably reflects the presence of nanoscopic
inhomogeneities intrinsic to the Ca-system which is beyond the detection limit of
EDX or we can speculate that the broad nature of superconducting transitions
is related to the extreme pressure sensitivity of this compound.

However, in order to obtain information about the electronic
properties in the superconducting state, we subtracted our 9 T specific heat data from the zero-field data for both
directions, labeled as \emph{$\Delta c_p/T$} in the insets of Fig. 2. The extracted superconducting transition temperature is found
to be 14.1 K, which is in agreement with resistivity and
magnetization studies on single crystals from the same batch
\cite{Harnagea2011},\cite{Klingeler2010}. Note, that this analysis for the estimation
of the transition temperature \emph{$T_c$} can be applied since the
superconducting transition of
Ca(Fe$_{0.937}$Co$_{0.063}$)$_2$As$_{2}$ is strongly suppressed
and shifted to lower temperaturesin in an external magetic field of 9T.
The same procedure has been used to extract \emph{$T_c$} for other
fields up to 6\,T (shown in Fig. 3) using the half jump height. From Fig. 3 it can be clearly
seen that the superconducting transition is systematically shifted
to lower temperature and reduced in height with increasing
magnetic field. This behavior occurs for both field directions.

From the field dependence of \emph{$T_c$} the phase diagram of
Ca(Fe$_{0.937}$Co$_{0.063}$)$_2$As$_{2}$ can be drawn, which is
shown in the lower inset of Fig. 3 depicting the experimental data
as well as the estimated upper critical field values $H_{c2}$ for
both directions. According to the Ginzburg-Landau (GL) equation~\cite{Woollam1974}, $H_{c2}$ can be expressed as
\begin{equation}\label{eq1}
    H_{c2}(t)=H_{c2}(0)[\frac{1-t^{2}}{1+t^{2}}],
\end{equation}
where \emph{$t$} is the reduced temperature \emph{$t = T/T_{c}$}. The upper
critical field at \emph{$T$} = 0 has been evaluated to be
\emph{$\mu$$_{0}$$H_{c2}^{(c)}(0)$} = 12.7 \,T and \emph{$\mu$$_{0}$$H_{c2}^{(ab)}(0)$} = 25.4 \,T for
the $\emph{c}$ and $\emph{ab}$ direction, respectively. Another possibility to extract the
upper critical field is to consider the single-band
Werthamer-Helfand-Hohenberg (WHH) model~\cite{Werthamer1966}:

\emph{H$_{c2}$(0)= - 0.69 $T_{c}(dH_{c2}/dT)_{T_{c}}$}. Our data are
perfectly described by a linear fit to the WHH model with an
average slope -\emph{$\mu$$_{0}$$dH_{c2}/dT$} = 1.2 \,T/K and 1.95 \,T/K
yielding upper critical fields $\mu$$_{0}$\emph{$H_{c2}(0)$} = 11.5\,T and 19.4\,T
for the $\emph{c}$ and $\emph{ab}$ direction, respectively. Note, that resistivity studies on CaFe$_{1.94}$Co$_{0.06}$As$_{2}$ single crystals show similar values for the average slopes, i.e., -\emph{$\mu$$_{0}$dH$_{c2}$$^{(c)}$/dT} = 1.4 T/K for B $\parallel$ \emph{c} and
-\emph{$\mu$$_{0}$dH$_{c2}$$^{(ab)}$/dT} = 1.8 T/K for B $\parallel$ \emph{ab}
\cite{Kumar2009}.

\begin{figure}[t,center]\vspace{-1.0pc}
\includegraphics[width=22pc,clip]{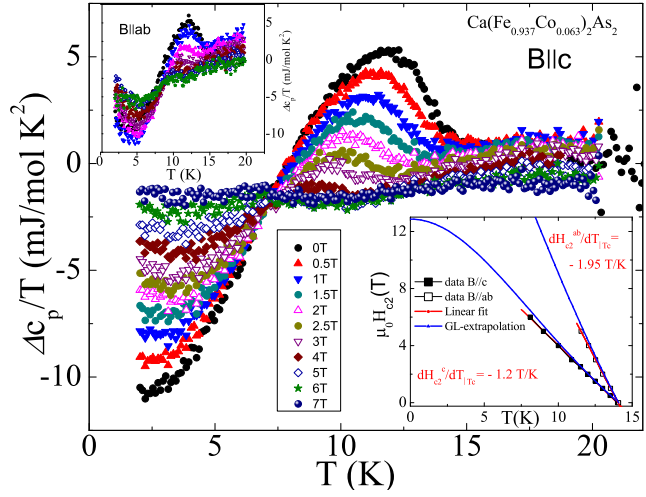}\hspace{2pc}%
\begin{minipage}[b]{15pc}\caption{\label{label} Specific heat data \emph{$\Delta c_p/T$} = \emph{$c_p (B) - c_p (9 T)$}
of Ca(Fe$_{0.937}$Co$_{0.063}$)$_{2}$As$_{2}$ in various magnetic
fields for $\emph{B} \parallel \emph{c}$. The upper inset shows \emph{$\Delta c_p/T$}
for $\emph{B} \parallel \emph{ab}$. The lower inset depicts the phase diagram
estimating the upper critical fields, where the blue dotted lines
show the theoretical curves based on the G-L theory and the red
dotted lines represent linear fits with average slopes of
-$\mu$$_{0}$\emph{$dH_{c2}^{(c)}/dT$} = 1.2 T/K and -$\mu$$_{0}$\emph{$dH_{c2}^{(ab)}/dT$} = 1.95 T/K
for the $\emph{c}$ and $\emph{ab}$ directions, respectively.} \label{8}
\end{minipage}
\end{figure}

\begin{figure}[t,center]\vspace{-1.0pc}
\includegraphics[width=22pc,clip]{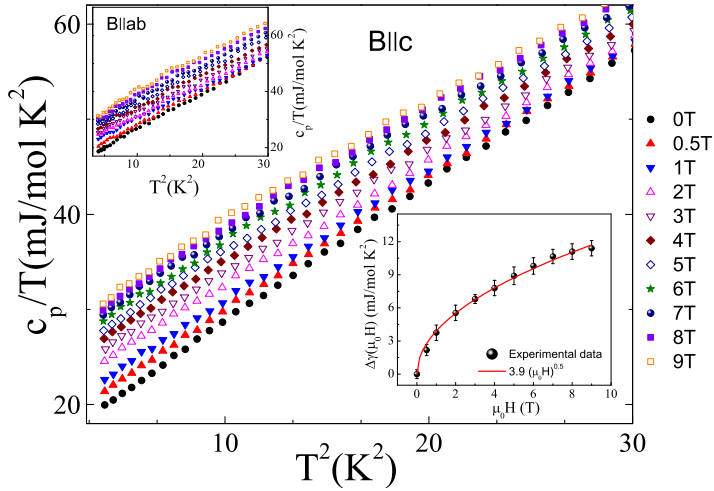}\hspace{2pc}%
\begin{minipage}[b]{15pc}\caption{\label{label}The electronic specific-heat coefficient \emph{$c_{p}/T$} vs. \emph{$T^{2}$} of Ca(Fe$_{0.937}$Co$_{0.063}$)$_2$As$_{2}$
for magnetic fields $\emph{B} \parallel \emph{c}$ up to 9\,T. The upper inset
shows the data for $\emph{B} \parallel \emph{ab}$. Both data show a rather
linear behavior and no Schottky anomaly is detected at low
temperatures. The lower inset depicts the field dependence of
\emph{$\Delta$$\gamma (H)$} for $\emph{B} \parallel \emph{c}$ . The solid red line represents a fit to \emph{$\Delta$$\gamma$(H)} =
\emph{A($\mu$$_{0}$H)$^{0.5}$}.}\label{8}
\end{minipage}
\end{figure}

Figure 4 depicts the low temperature specific heat of
Ca(Fe$_{0.937}$Co$_{0.063}$)$_2$As$_{2}$ plotted as \emph{$c_{p}/T$} vs.
\emph{$T^{2}$} in different applied magnetic fields for $\emph{B}
\parallel \emph{c}$. In the upper inset the experimental data for $\emph{B} \parallel \emph{ab}$ are depicted.
One can see the roughly linear behavior for both directions and
the absence of magnetic impurities in our sample, as usually
evidenced by a Schottky anomaly at low temperatures. It is clear
that the magnetic field enhances the low-temperature specific heat
continuously, indicating the increase of the quasiparticle
density-of-states at the Fermi level induced by a magnetic field.
A linear extrapolation of the low-T data to zero temperature
yields the field dependence of the field-induced contribution
\emph{$\Delta\gamma(H)$ = $(C(T,H) - C(T,0))/T$} at \emph{$T$} = 0. The lower
inset of Fig. 4 shows \emph{$\Delta\gamma$(H)} for $\emph{B} \parallel \emph{c}$,
clearly indicating a non-linear behavior in magnetic fields. In
fact, this non-linear behavior can be fitted by the simple
equation \emph{$\Delta$$\gamma$(H)} =
\emph{A($\mu$$_{0}$H)$^{0.5}$} predicted for d-wave
symmetry in the clean limit \cite{Volovik1993}, matching our
experimental data. In Ba(Fe$_{1-\emph{x}}$Co$_{\emph{x}}$)$_{2}$As$_{2}$
single crystal at (\emph{x} = 0.045, 0.08, 0.103, and 0.105), the \emph{$\gamma$(H)} values of the
low-temperature specific heat data also fit well within this clean-limit d-wave
model \cite{Gofryk2010}. However, it is difficult to solely judge
from specific heat data wether nodes exist or not and it should be noted that
other pairing mechanisms can also result in a non-linear behavior
of \emph{$\gamma(H)$}.

\section{Conclusions}

In summary, we have studied the heat capacity of single crystals
of Ca(Fe$_{1-\emph{x}}$Co$_{\emph{x}}$)$_{2}$As$_{2}$ (\emph{x} = 0, 0.032, 0.051, 0.056,
0.063, and 0.146). Here, a combined magnetic and structural
ordering has been observed for the parent compound evidenced by a
first order transition around 168\,K. For higher doping levels
this transition shifts to lower temperatures, until it is
completely suppressed for x = 0.056 and superconductivity appears
around 14.1 K. From the low temperature specific heat data of the Co-doped superconducting
Ca(Fe$_{0.937}$Co$_{0.063}$)$_{2}$As$_{2}$ single crystal, the
upper critical field has been determined using the WHH model. It
is found to be 11.5\,T and 19.4\,T for $B
\parallel c$ and $\emph{B} \parallel \emph{ab}$, respectively. A non-linear behavior of the
field dependence of \emph{$\gamma (H)$} in the low-temperature region hints towards d-wave superconductivity in this compound.

\section*{Acknowledgments}
We thank M. Deutschmann, S. M\"{u}ller-Litvanyi, R. M\"{u}ller, J.
Werner, S. Pichl, K. Leger and S. Gass for technical support. This
project was supported by the DFG through BE1749/13.

\end{document}